**WISE/NEOWISE Preliminary Analysis and Highlights of the 67P/Churyumov-Gerasimenko Near Nucleus Environs**

James M. Bauer[1,2], Emily Kramer[1,3], A. K. Mainzer[1], Rachel Stevenson[1], Tommy Grav[4], Joseph R. Masiero[1], Russell G. Walker[5], Yan R. Fernández[3], Karen J. Meech[6,7], Carey M. Lisse[8], Paul R. Weissman[1], Roc M. Cutri[2], John W. Dailey[2], Frank J. Masci[2], David J. Tholen[7], George Pearman[2], Edward L. Wright[9], and the WISE Team

[1]Jet Propulsion Laboratory, California Institute of Technology, 4800 Oak Grove Drive, MS 183-401, Pasadena, CA 91109 (email: bauer@scn.jpl.nasa.gov)

[2]Infrared Processing and Analysis Center, California Institute of Technology, Pasadena, CA 91125

[3]Department of Physics, University of Central Florida, 4000 Central Florida Blvd., P.S. Building, Orlando, FL 32816-2385

[4]Planetary Science Institute, 1700 East Fort Lowell, Suite 106, Tucson, AZ 85719-2395

[5]Monterey Institute for Research in Astronomy, 200 Eighth Street, Marina, CA 93933

[6]Institute for Astronomy, University of Hawaii, 2680 Woodlawn Dr., Manoa, HI 96822

[7]NASA Astrobiology Institute, University of Hawaii, Manoa, HI 96822

[8]Applied Physics Laboratory, Johns Hopkins University, 11100 Johns Hopkins Road Laurel, MD 20723-6099

[9]Department of Physics and Astronomy, University of California, PO Box 91547, Los Angeles, CA 90095-1547

**ABSTRACT**

On January 18-19 and June 28-29 of 2010, the Wide-field Infrared Survey Explorer (WISE) spacecraft imaged the Rosetta mission target, comet 67P/Churyumov-Gerasimenko. We present a preliminary analysis of the images, which provide a characterization of the dust environment at heliocentric distances similar to those planned for the initial spacecraft encounter, but on the outbound leg of its orbit rather than the inbound. Broad-band photometry yields low levels of $CO_2$ production at a comet heliocentric distance of 3.32 AU and no detectable production at 4.18 AU. We find that at these heliocentric distances, large dust grains with mean grain diameters on the order of a millimeter or greater dominate the coma and evolve to populate the tail. This is further supported by broad-band photometry centered on the nucleus, which yield an estimated differential dust particle size distribution with a power law relation that is considerably shallower than average. We set a 3σ upper limit constraint on the albedo of the large-grain dust at ≤ 0.12. Our best estimate of the nucleus radius (1.82 +/- 0.20 km) and albedo (0.04 +/- 0.01) are in agreement with measurements previously reported in the literature.

## 1. Introduction

The Rosetta mission primary target, comet 67P/Churyumov-Gerasimenko (hereafter 67P), has been extensively studied at visual and infrared wavelengths. The spacecraft will arrive at 67P in June, 2014 when the comet is 4.0 AU from the Sun, inbound and moving toward its perihelion at q= 1.24 AU on August 15, 2015. After orbiting the nucleus for several months, the spacecraft will release the Philae lander in November, 2014 toward the surface, while the mother spacecraft follows the comet as it passes through perihelion and out to a heliocentric distance of more than 1.8 AU (cf. http://www.esa.int/esaMI/Rosetta/ESAIBF7708D_0.html ). Successful performance of the nominal Rosetta mission requires a high-fidelity understanding of the coma environment of the comet. The salient features previously reported in the literature for 67P include nucleus radius estimates ranging from ~1.7-2.6 km yielding surface albedo values ranging from 0.034-0.043 with a 12.4-12.7 hour rotation period (Lamy *et al.* 2008, Kelley *et al.* 2009), and a dust trail composed of particles in excess of 100μm in size (Ishiguro 2008, Kelley *et al.* 2008). The infrared measurements (Kelley et al. 2008) were taken in 2004, when the comet was inbound at heliocentric distances in the range of 4-4.5 AU, i.e. at similar distances to when the Rosetta spacecraft will enter orbit around 67P.

The WISE spacecraft (Wright et al. 2010) observed 67P twice over the course of its cryogenic mission, when the comet was outbound at distances of 3.32 and 4.18 AU from the Sun. These data present the opportunity to study the large-grain dust coma and trail of the comet at different epochs, and compare the behavior of the comet across different orbits. In this paper we present a preliminary analysis of the WISE imaging photometric data.

The WISE mission surveyed the sky at four mid-IR wavelengths simultaneously, 3.4 μm (W1), 4.6 μm (W2), 12 μm (W3) and 22 μm (W4), with approximately one hundred times improved sensitivity over the Infrared Astronomical Satellite (IRAS) mission (Wright *et al.* 2010). The field of view for each exposure was 47 arcmin on a side, and an individual square pixel spanned 2.8 arcseconds. Over 99% of the sky was covered with multiple exposures, averaging 10 per sky region, but varying in density as a function of ecliptic latitude (Cutri *et al.* 2012), with the largest number of exposures occurring at the ecliptic poles and the least at the ecliptic equator. On the ecliptic, the average number of exposures per sky region was 10-12. For this paper we have utilized an enhancement to the WISE data processing system called "NEOWISE". The WISE Moving Object Processing Software (WMOPS) was developed to find solar system bodies in the WISE images (Dailey *et al.* 2010, Mainzer *et al.* 2011a). WMOPS successfully found a wide array of primitive bodies, including Near-Earth Objects (NEOs), main belt asteroids, comets, and Trojan and Centaur asteroids. By the end of the spacecraft mission, NEOWISE identified more than 157,000 small bodies, including 150 comets (cf. Mainzer *et al.* 2011a, Bauer *et al.* 2012). These infrared observations are useful for determining solid body size and albedo distributions, and thermo-physical properties such as thermal inertia, the magnitude of non-gravitational forces, and surface roughness (Mainzer *et al.* 2011b, 2011c). The subset of these bodies exhibiting cometary activity require special treatment to interpret the observations. The observed fluxes and morphologies are significantly affected by the coma material surrounding the solid nucleus, i.e. by the contribution to the IR flux from the gas and dust, and the variable nature of the observed brightness of the object attributable to changing heliocentric distance and/or outbursts. These IR imaging data provide unique opportunities to characterize four main components of comets: the nucleus, the dust and gas comae, and the extended dust trail. Here we apply specialized techniques for the study of comets in the WISE data described in Bauer *et al.* (2011, 2012) to the primary target of the Rosetta mission, comet 67P/Churyumov-Gerasimenko, with emphasis on the region in close proximity to the nucleus. A more extensive analysis of the dust trail and other aspects will be discussed in a later work.

## 2. Observations & Analysis

The WISE spacecraft surveyed the sky as its terminator-following geocentric polar orbit around the Earth progressed about 1 degree of ecliptic longitude per day. Regular survey operations commenced on Jan. 14, 2010 (MJD 55210), imaging the sky simultaneously in all four bands until the solid hydrogen cryogen was depleted in the secondary tank on Aug. 5, 2010 (MJD 55413). The survey then entered a three-band (W1-W3) phase that lasted through Sep. 30, 2010 (MJD 55469). The final phase, the post-cryogenic mission with only W1 and W2 operating, lasted from Oct. 1, 2010 through Jan, 31, 2011 (MJD 55592; cf. Cutri *et al.* 2011). All photometric data of detected objects presented here were obtained during the cryogenic phase.

During the fully cryogenic portion of the mission, simultaneous exposures in the four WISE wavelength bands were taken once every 11 sec, with exposure durations of 8.8 sec in W3 and W4, and 7.7 sec in W1 and W2 (Wright *et al.* 2010). The number of exposures acquired on each moving object varied depending on the location of the object on the sky, especially its ecliptic latitude, the toggle pattern of the survey employed to avoid imaging the Moon, and the relative motion of the object with respect to the progression of the survey (Mainzer *et al.* 2011a, Cutri *et al.* 2011). Note that WISE may have observed an object while it was in different patches of the sky, i.e. when several weeks or months had passed since the previous exposure; henceforth, we refer to the series of exposures containing the comet in the same region of sky as a "visit".

The spatial resolution in the WISE images varies with the wavelength of the band. The full-width-half-maximum (FWHM) of the mean point-spread-function (PSF), in units of arcseconds is 6.1, 6.4, 6.5 and 12.0 for W1, W2, W3, and W4, respectively (Wright *et al.* 2010). Table 1 summarizes the observing circumstances for 67P. Some observations were near the edge of the imaging area or near a chip artifact. We excluded these observations from the image stack and light curve photometry. A total of 11 observations were taken during the first visit (hereafter Visit A, or VA) and 15 during the second (hereafter Visit B or VB). The number of frames included in the analyses of each visit was 10 and 11, respectively for VA and VB. Both visits occurred post-perihelion, with the comet at heliocentric distances similar to those during the Spitzer Space Telescope observations reported by Kelley *et al.* 2008 and Lamy *et al.* 2008. The Spitzer observations were taken pre-perihelion when the comet was inbound, however, and so provide a unique point for comparison with the WISE data.

 The WISE image data were processed using the "first pass" scan/frame pipeline that applied instrumental, photometric, and astrometric calibrations (Cutri *et al.* 2011). WISE covered all ecliptic latitudes each day in two narrow longitude wedges at 95 +/- 2 degrees ahead of the Sun and 90 +/- 2 degrees behind the Sun.  It used the Earth's orbital motion around the Sun to scan these wedges across all ecliptic longitudes over 6 months. The comet's maximum rate of angular motion on the sky at the time of the observations was less than 20 arcseconds per hour. Thus, the apparent motion created a blurring of at most ~0.05 arcsec, an insignificant factor in the imaging, as the blur was far smaller than the pixel scale at the shorter wavelengths (2.75 arcsec/pixel in W1, W2, and W3; 5.5 arcsec/pixel in 2×2-binned W4; Wright *et al.* 2010).

The images from each visit were stacked to increase signal from the comet. The images were shifted to match the sky-motion rates of 67P as predicted by JPL's Horizon's ephemeris service (http://ssd.jpl.nasa.gov). The images were co-added using the "A WISE Astronomical Image Co-adder" (AWAIC) algorithm as described in Masci & Fowler (2009), which produces a re-sampled image with a 1 arcsec per pixel scale. The images were stacked in this manner for each corresponding visit to the comet to conduct the photometric and morphological analyses.

**Table 1: Mid-IR Observations of 67P Summary**

| Object[a] | N[a] | R [a] (AU) | Δ [a] (AU) | α[a] (°) | Comments |
|---|---|---|---|---|---|
| 67P, VA | 11 | 3.32 | 3.31 | 17.2 | Strong detections in W3 & W4, weak in W1 &W2 |
| UT[a] Start Times: 2010-Jan-18 19:10:40.722, 2010-Jan-18 22:21:11.852, 2010-Jan-18 22:21:22.852[*], 2010-Jan-19 01:31:53.982, 2010-Jan-19 03:07:04.045, 2010-Jan-19 04:42:25.104, 2010-Jan-19 06:17:46.172, 2010-Jan-19 07:52:56.235, 2010-Jan-19 09:28:17.302, 2010-Jan-19 12:38:48.428, 2010-Jan-19 15:49:19.558 | | | | | |
| 67P, VB | 15 | 4.18 | 3.98 | 14.0 | Strong detections in W3 & W4, none in W1 & W2 |
| UT Start Times: 2010-Jun-28 20:41:10.530, 2010-Jun-28 23:51:41.577, 2010-Jun-29 03:02:01.623, 2010-Jun-29 03:02:12.628, 2010-Jun-29 06:12:32.662, 2010-Jun-29 06:12:43.666[*], 2010-Jun-29 07:47:53.685, 2010-Jun-29 09:23:03.708 , 2010-Jun-29 10:58:24.731, 2010-Jun-29 12:33:34.751, 2010-Jun-29 14:08:55.774[*], 2010-Jun-29 17:19:26.816, 2010-Jun-29 20:29:57.863[*], 2010-Jun-29 23:40:17.909[*], 2010-Jun-29 23:40:28.909 | | | | | |

*2.1 Nucleus –*

In order to extract the nucleus signal, we adapted routines developed by our team (Lisse *et al.* 1999, 2009, Fernández 1999, and Fernández *et al.* 2000, 2012) to fit the coma as a function of angular distance from the central brightness peak along separate azimuths, as done in Bauer *et al.* 2011. As per the description in Lisse et al. 1999, the model dust coma was created using the functional form $f(\Theta) * \rho^{-n}$, where $\rho$ is the projected distance on the sky from the nucleus, and $\Theta$ is the azimuthal angle. In order to compensate for the WISE instrumental effects, the model coma was then convolved with the instrumental PSF appropriate for AWAIC co-added images (see Cutri et al. 2012). Radial cuts through an image of the comet were made every 3° in azimuth, and the best-fit radial index, n, and scale, f, at each specific azimuth were found by a least-squares minimization of the model to the data along that azimuth. The pixels between 5 and 20 arcsec of the brightness peak were used to fit the model coma.  The coma model fit residuals were $\sim$10% for the W3 and W4 images, similar to the photometric uncertainties in the nucleus and coma signals.  We use the nucleus mean values of absolute magnitudes from Hubble Space Telescope observations reported in Lamy *et al.* (2006), i.e. $H_V$=16.3 for our nucleus albedo calculations, as these constitute the highest spatial-resolution nucleus measurements in the literature. Note that Tubiana *et al.* (2008) report $H_V$=15.9, which would raise our derived albedos from VA and VB by factors of ~1.4. We performed the extractions independently for the two visits, using the stacked images (Figure 1) to both average over the rotational variations and boost the signal-to-noise ratio (SNR). The higher SNR visit (see Figure 2), VA, had extracted nucleus signals 1.9 +/- 0.17 mJy and 5.9 +/- 0.9 mJy in W3 and W4, respectively, and were fit to a NEATM model (Harris 1998, Delbo *et al.* 2003, Mueller et al. 2009 and Mainzer *et al.* 2011b) with a free beaming ($\eta$) parameter, yielding a diameter of 3.64 +/- 0.39 km, an albedo (herafter $p_v$) of 0.04 +/- 0.01, and an $\eta$ value of 1.02 +/- 0.17. The $\eta$ value obtained in the free fit is similar to the mean value of $\eta$ =1.03 +/- 0.11 found for 55 comets out of the sample obtained by Fernández *et al.* 2012 from a Spitzer survey of 100 Jupiter Family Comets. We obtained lower signal-to-noise extractions for VB (the comet was at a larger heliocentric distance during that visit), with W3 and W4 flux values of 0.28 +/- 0.04 and 3.6+/- 0.8 mJy, respectively, yielding a diameter of 3.0 +/- 1.4 km and $p_v$ = 0.06 +/- 0.03 for a fixed $\eta$ value of 1.0. We held $\eta$ fixed for the VB thermal fit since the free fit solution converged to a non-physical values of > $\pi$ (Harris 1998). The diameter values are consistent with the average spherical radius value obtained by Kelley *et al.* (2008; 1.87 +/- 0.08 km) and Lamy *et al.* (2008; 1.98 +/- 0.05 km). A nucleus of this size will produce combined reflectance and thermal flux contributions of 3$\mu$Jy and 6$\mu$Jy in W1 and W2, respectively, well below the uncertainty of any flux or flux upper limits in those bands. Still, the flux contributions from the nucleus were subtracted from the flux values of the coma reported in Table 2.

*2.2 Dust Photometry & CO/CO$_2$ production –*

Lisse *et al.* 1998 demonstrated how broad-band photometry can be applied to determine the quantity and temperature of the coma dust. As in Bauer *et al.* 2011 & 2012, we performed blackbody temperature fits to the dust coma region surrounding the nucleus, extracting the W3 and W4 measured nucleus flux contribution from the thermal signal. We also estimated the W1 and W2 nucleus contribution, based on the thermal fit results for the size and albedo computed using W3 and W4 and scaling the reflected light contribution to the appropriate phase angle, assuming an IR albedo twice that of the visual-wavelength albedo (Grav et al. 2011). W3 and W4 fluxes from the nucleus represent less than 11% of the total signal in the 11-arcsec aperture photometry for VA. The thermal fit was dominated by the dust coma to within the photometric uncertainty, and the uncertainty in the removed nucleus contribution was much less than the photometric uncertainty in the coma signal. For VB, the extracted nucleus fluxes comprise between 10% (W3) and 20% (W4) of the thermal flux in the 11 arcsecond aperture. The associated uncertainties were still less than the aperture flux uncertainties, but are still included in the flux uncertainties listed in Table 2. The results of the dust thermal fits are listed in Table 2, and the dust fits for each visit are shown in Figures 2A & 2B. The signal contribution from the nucleus, as estimated from the thermal fit results, was removed from the total flux in the derivation of the coma flux. We calculate the effective area for the dust using 9 and 11 arcsecond radius apertures, and find that the effective area matches across both thermal bands, W3 and W4. In fact, for the thermal bands, this derived area has a factor of the emissivity, $\varepsilon$, incorporated into the result. Division by the projected length scale of the apertures, i.e. the $\rho$ value, and by the constant $\pi$, provides an aperture-independent means of comparison to the quantity of dust visible at particular wavelengths analogous to $Af\rho$ (A'hearn et al. 1984). We call this factor $\varepsilon f\rho$, as introduced by Lisse *et al.* (2002) and used in Kelley *et al.* (2012), which is listed in Table 2. The value of $\varepsilon f\rho$ assumes the observed flux is attributable primarily to the dust continuum emission and is the product of the emissivity, $\varepsilon$, the fractional area within an aperture filled by the dust, *f*, and the projected length scale of the aperture radius on the sky at the distance of the comet, $\rho$, expressed in cenitmeters. We compute our $\varepsilon f\rho$ values multiplying the observed surface brightness of the comet, $I_\nu$, by $\rho$ and dividing by the Planck function, $B_\nu(T_b)$ where $T_b$ is the *fitted* black-body temperature of the dust, here 183K for VA and 169K at VB. The effective area is derived from this quantity by multiplying by $\pi\rho/\varepsilon$ using and assumed value of $\varepsilon\approx0.9$. Derived values for dust temperature, $Af\rho$, $\varepsilon f\rho$, $Q_{CO2}$ and $Q_{CO}$, from the 9 and 11 arcsecond apertures match within the photometric uncertainty of the 11 arcsecond coma flux signal (listed in Table 2).

**Table 2: Coma Photometry Results**

| Visit | Coma Flux[a] | | | | $T_{dust}$[a] (K) | W1 $Af\rho$[a] (cm) | W3 $\varepsilon f\rho$[a] (cm) | W4 $\varepsilon f\rho$[a] (cm) | $Q_{CO2}/Q_{CO}$[a] (s$^{-1}$) |
|---|---|---|---|---|---|---|---|---|---|
| | W1 ($\mu$Jy) | W2 ($\mu$Jy) | W3 (mJy) | W4 (mJy) | | | | | |
| VA | 10±7 | 180±50 | 10.2±.4 | 50±4 | 183±4 | 8.7±4 | 146±24 | 147±24 | $7.7(\pm2)\times10^{25}$ $8.2(\pm2)\times10^{26}$ |
| VB | -- | -- | 2.8±.2 | 19±2 | 169±3 | -- | 82±10 | 90±21 | $<10^{26}, <10^{27}$ |

[a] Each coadded photometry point is listed for each visit (see text). The fluxes here, in units of milli-Janskys for W3 and W4, and micro-Janskys for W1 and W2, and are reported for 11 arcsecond aperture radius values, with the nucleus contribution to the signal removed. $T_{dust}$ is derived from the black body fits to the coma flux values, where the expected black-body temperatures are 157 and 140K for VA and VB respectively. $Af\rho$ is calculated from the 1.5 SNR signal in W1, and $\varepsilon f\rho$ from the W3 and W4 coma fluxes, and are expressed in [cm] units along with the uncertainties. $Q_{co2}$ and $Q_{co}$ are derived from the W2 excess, assuming only a single species, either $CO_2$ or $CO$, is accountable for the entire excess signal.

Thermal fits of the dust to a black body were made using the W3 and W4 fluxes for VA and VB, and were found to exceed the expected black body temperature by ~20%. This is not uncommon (cf. Lisse et al. 1998), but is usually seen for dust size distributions with a presence of small grains. Note that at the infrared wavelengths used by WISE, the observations are not sensitive to fine cometary dust on the order of 1 μm, which typically dominates the visible signal for coma photometry.

For VA, we find a W2 excess signal from the dust at the 2σ level. Scaling the dust black body curve and expected reflectance given a standard α=-3 power law for the particle size distribution (PSD; cf. Fulle 2004), with a neutral reflectance and a reflectance reddening law based on Jewitt & Meech (1986) averaged out to 3.5μm, we employ the method from Bauer *et al.* 2011 to determine the excess flux. This is somewhat conservative in its approach for this particular comet, as the PSD is quite flattened relative to the standard PSD for 67P at these distances (see below). We emphasize that the detection is only 2σ above the dust flux estimates for W2. Using the methods outlined by Pittichová *et al.* (2008) and applied in the case of the WISE filter bands in Bauer *et al.* (2011, 2012), we find the production rate estimates listed in Table 2. We find no excess in the VB signal, and indeed no significant W2 signal at that time. The 1σ upper limits listed in Table 2 provide a relatively loose constraint on production during VB owing to the comet's heliocentric distance of 4.18 AU. The W1 signal listed in Table 2 is of an even lower SNR of 1.4. Still, we have calculated a corresponding A$f\rho$ value, and we use it later to constrain the dust properties, assuming any signal is primarily from the dust's reflected light.

*2.3 Dust Morphology –*

Figures 1C and 1D show the dust morphology in the highest resolution thermal band, W3. Compared to the color Figures 1A and 1B, the morphology is similar in both thermal bands. Finson-Probstein dust models (Finson & Probstein 1968) for various $\beta$ parameters were performed using the comet viewing geometries at the time of VA and VB. The $\beta$ parameter describes the ratio of the force on dust grains attributed to solar radiation pressure relative to solar gravity. In physical units, this gives the ratio (Burns et al. 1979; Finson & Probstein 1968):

$$\beta = 1.19 \times 10^{-4} \frac{Q_{PR}}{\rho_d d} \tag{1a}$$

where $\rho_d$ is the dust grain density in g cm$^{-3}$, d is the grain diameter in cm, and $Q_{pr}$ is the scattering efficiency for radiation pressure, with $Q_{pr} \sim 1$ for grain sizes of d>$\lambda$ (Burns *et al.* 1979), where $\lambda$ is the wavelength of the observation. As the WISE bands are relatively insensitive to grains with d ≤ $\lambda$, particularly with respect to the peak wavelength of scattered light from the Sun, this relationship applies. Larger $\beta$ (≥ 0.1) implies a stronger coupling to solar radiation pressure. Assuming a grain density ~1 g cm$^{-3}$, $\beta$ values ~0.01 correspond to grain sizes of ~100 $\mu$m. The value of $\beta$ is incorporated into the equation of motion in the following way:

$$\ddot{\vec{x}} + (1 - \beta) \frac{GM_s}{|\vec{x}|^3} \vec{x} = 0 \qquad (1b)$$

where $G$ is the universal gravitational constant, $M_s$ is the mass of the Sun, and $\vec{x}$ is the vector position of the object. This is a simple equation of motion that can then be integrated for different values of $\beta$ to track the motion of particles with a particular $\beta$ value.

The computations used a numerical integrator written in *Python* (based on the work of Lisse et al., 1998) which took a set of $\beta$ values, integrated the motion of the dust particles over the designated time interval, and returned the syndyne for each $\beta$ input. A syndyne is the path of dust grains released from the nucleus of a comet continuously with a particular value of $\beta$. This created a set of curves (syndynes) that show the theoretical positions of dust with a particular $\beta$-value that was released starting at some initial time up to the time the image data were observed. Since the forces on particles of different $\beta$ are relatively different, the syndynes will tend to fan out in the comet's orbital plane. If the data are truly represented by the syndynes, the curves will span the width of the dust tail when overplotted on the comet image. In the general case, the relative velocity of the grains from the comet's surface can be included into the integration, giving the curves a spread of some finite width. In this case, we are interested in matching the large-scale morphology, so the particles were given no initial velocity. The syndynes that best matched the data, for both visits, were two years old, 8 to 13 months before its February 28, 2009 perihelion, and for $\beta \sim 0.001$, corresponding to dust grain diameters $\sim$1mm.

*2.4 Light Curve Analysis –*

We extracted thermal signal in the highest-resolution thermal band, W3, using a 9-arcsecond aperture radius surrounding the central condensation of the comet (see Figure 2C). This aperture size corresponded to 3 times the W3 PSF FWHM (Cutri *et al.* 2012). After correcting for flux offsets between visits, the data were phased to a period of 12.65 hours (Lamy *et al.* 2008). The data were too sparsely sampled to confirm or refute the period. However, the data are consistent with the low-amplitude ($\Delta$m = 0.4 +/- 0.1) of brightness variation reported previously (Kelley et al. 2008), and implying a minimum *a/b* principal axis ratio of 1.45 +/- 0.15 (cf. Meech *et al.* 1997).

## 3. Discussion

*3.1 Dust Production –* We calculated the $\epsilon f \rho$ for the dust, and found that the values matched across both thermal bands and apertures within the uncertainties. Converting these values into effective particle column densities leaves no significant increase in the number of grains as a function of decreasing wavelength, implying a discrete particle size distribution slope much less than $\alpha$=-3 (cf. Fulle *et al.* 2004). In fact, we find between sizes of 11 $\mu$m and 22 $\mu$m, there is no excess area, implying that most of the dust in the aperture is of size 22 $\mu$m or larger, i.e. that the *cumulative* dust particle size distribution is flat below 22 $\mu$m. This is likely caused by solar radiation pressure removing smaller grains from an old dust population, as significant dust emission has ceased some months before.

The dust trail analysis (Figure 1) in combination with the dust coma photometry places additional constraints on dust production. Dust production is better constrained using IR observations as compared to optical alone, since most of the mass resides in larger grains, while the optical signal is dominated by smaller grains (cf. Bauer *et al.* 2008). Assuming dust densities ~1 g cm$^{-3}$, we derive dust sizes ~ 1mm for the $\beta$~0.001 syndyne (see eq. 1a and Figures 1C & 1D), the larger of the $\beta$ values with orbits that possibly match the morphology. This implies an individual grain mass of $5\times10^{-4}$ grams. By multiplying our value of $\varepsilon f\rho$ by $\pi\rho$ and dividing by an assumed emissivity value $\varepsilon$~0.9, we can derive an effective area. Our $\varepsilon f\rho$ value derived from the 22 $\mu$m flux yields an effective total area of $1.4\times10^{12}$ cm$^2$, and with a mean individual grain area of $7.85\times10^{-3}$ cm$^2$, yields a grain count of $N_{dust}$~$1.8\times10^{14}$ grains for the 11 arcsecond aperture. This yields an average dust grain column density of $6 \times 10^{-6}$ grains per cm$^2$ within ~32000 km of the nucleus, or an average density of ~$2 \times 10^{-9}$ grains per m$^3$. If we use the maximum grain size from Ishiguro et al. (2008), we still derive $6 \times10^{12}$ grains within the 11 arcsecond aperture and a corresponding $8 \times 10^{-11}$ grains per m$^3$, though the morphology suggests these will be concentrated in the dust tail.   With a grain density of 1 g cm$^{-3}$, we find a total mass of $9\times10^{10}$ grams within the 11 arcsecond aperture for VA, similar to other JFC coma dust mass estimates (cf. Lisse et al. 2002). Scaling the length of the $\beta$~0.001 syndyne, a mean crossing time for the 11 arcsecond aperture is ~12+/-3 days, yielding a mass loss rate ~7 (+/- 2) $\times 10^4$ g/s. Similar analysis of the VB data yields a comparable mass loss rate ~4(+/- 1) $\times 10^4$ g/s, i.e. in agreement with VA within the uncertainties of crossing time and measured flux. At the calculated mass loss for VA, the orbital mass loss is 1 to 1.5× $10^{13}$ g/orbit, or on the order of other JFCs of similar size (cf. Reach *et al.* 2000, Lisse et al. 2002) which range between $\sim 10^{12}$ and $10^{15}$ g/orbit. Note, however, that this is a lower limit, as the mean particle size may be larger than the 1mm assumed from the morphological analysis of the dust. Also, these are based on only two spans of observations while the comet is outbound and at heliocentric distances where the water sublimation rate is low.

*3.2 Dust Reflectivity* - Reflected light signal is detected in W1 for only VA, and the detection is at the $\sim 1.4$-$\sigma$ detection strength (9.0 +/- 6.3 μJy for the coma flux, i.e. with the estimated nucleus signal removed). The W1 single VA SNR$\sim 1$ detection presents an opportunity to constrain dust production and PSD, but it also provides a further constraint on the dust grain albedo at 3.4 μm (hereafter $p_{W1}$). Since a higher reflectivity will result in a brighter signal at shorter, i.e. reflected-light, wavelengths from the quantity of dust observed at thermal wavelengths, we can actually constrain the albedo. Smaller dust grains scatter light more efficiently per unit mass. A quantity of dust with the same effective area in all bands implies a PSD $\alpha << 3$, where the number of dust grains of a size d, $N_{dust}(d)$, is proportional to d to the power $-\alpha$. Fulle *et al.* (2004) found values of $\alpha$ between 3 and 3.5 for 67P at smaller heliocentric distances out to 2.7AU. The effective area of emitted light from the larger-grains must not exceed the effective area derived for the reflected light of the same dust observed at shorter wavelengths. For grain reflectivity $\sim 0.03$, the effective areas match, leaving, as we saw, considerably fewer small-grains to account for the fraction of an effective area left when subtracting out the contribution from grains with sizes at or in excess of ~22 μm. In other words, the W1 signal is consistent with the quantity of dust seen in W3 and W4 if only 3% of the light is reflected. A more direct comparison is made by comparing the thermal band ε$f\rho$ and W1 reflected light $Af\rho$ values. Both should describe the same quantity of dust. Within the uncertainties, assuming an emissivity near 0.9,

$$p_{W1} \approx 0.9 \frac{Af\rho}{\varepsilon f\rho} \qquad (2)$$

which yields a value of $p_{W1} \approx 0.053$ +/- 0.04. However, with such a low SNR detection, it is not unambiguously caused by noise. In such a case, the 3σ detection limit would still place the $p_{W1} \leq 0.12$. To summarize, assuming the large-grain dust dominates both the reflected light at 3.4 μm (W1) and emitted light (W3 and W4), we find a mean grain $p_{W1}$ of 0.05 +/- 0.04, or alternatively place a 3-σ constraint on the large-grain $p_{W1}$ value of ≤ 0.12; in any case, the large-grain reflectance is not consistent with bright ice-dominated grains.

If we assume a significant small-grained (a few microns or less in size) contribution to the coma brightness, such that the contribution of the small grains to the W3 and W4 signal was not significant, while the contribution to the W1 signal was significant, the resultant mean albedo would be less than 0.05, assuming small grains and large grains shared the same reflectance. If the grains were very dark at 3.4 μm, i.e. if $p_{W1}$~0.02, the small grains would account for 1.6 times the effective area accounting for the signal at W3 and W4. Assuming further that the majority of these small grains had diameters ~3.4 μm, the derived value for the discrete size distribution α would be ∼2 to 2.3, considerably lower than the α = 3 to 3.5 reported by Fulle et al. (2004). Finally, though it may very well be the case that α ∼ 3 to 3.5 while 67P is more active, given the nearly equal ε$f\rho$ values for W3 and W4, it is not likely a power law adequately describes the size distribution of dust if small grain particles are present at these distances while the comet is outbound. Such a bi-modal dust particle size distribution may be possible if the large-grain dust, relatively unaffected by solar radiation pressure, is from a remnant trail, lingering from earlier activity, while more recent activity is weak, ejecting relatively few small dust grains from the surface.

Constraining the optical brightness without simultaneous optical imaging is complicated by the possibility of outburst during independent wavelength observations. Using the Minor Planet Center (MPC) database (http://www.minorplanetcenter.net) we find reported brightness values (of unspecified band-passes) between 18.4 and 20.4 (MPC 69186) spanning the time from Nov. 22, 2009 through March 10, 2010. Similarly, reported brightness values range from 19.5 to 20.5 (MPC 71054) at the time of the June observations. Because the aperture size and accuracy of the reported photometry are unknown, one cannot use these as constraints to our particle size distribution, as applied in Bauer *et al.* (2008), with the reported optical observations constraining particle sizes on the order of 0.5 to 1 micron. However, assuming an R-band brightness of ∼ 19 (+/- 1) magnitudes (i.e. near the 18.4 magnitude measurement corrected for phase and distance for VA) corresponds to the same aperture value as in VA, we derive an A$f\rho$ value of 14.2 (+21/-8) (in cm units), and find a $p_{vis} \sim 0.09$ (+0.10/-0.06) for the quantity of dust detected in the thermal bands, roughly consistent with our value for $p_{W1}$. Similarly, for VB, assuming an R-band magnitude of ~20, $p_{vis} \sim 0.06$. As the presence of some small-grained dust cannot be completely ruled out, owing to the temperature excess (Section 2.2), these reflectances based on the reported visual band magnitudes should be considered as crude upper limits to the large-grain dust albedo.

*3.3 CO2 Activity* - It is important to note the similarities and differences between the observed behavior of 103P/Hartley 2 and 67P. We find in both comets evidence for a persistent trail, and clear evidence of large-grained particles at heliocentric distances far from each comet's perihelion (cf. Bauer *et al.* 2011). The $CO_2$ detection in 103P was clearly more pronounced, though it was observed at a considerably closer heliocentric distance, and evidence reported here exists for some $CO_2$ activity in 67P. Other comets show activity at large distances beyond 3 AU (cf. Stansberry *et al.* 2003, Lisse *et al.* 2004, Meech et al. 2009). Thus such behavior is not unprecedented. However, we might expect to find more evidence of the existence of small-grained dust than we do here, resulting in a more significant W1 flux. It must be emphasized that the W2 excess signal is only at the ~2 to $3\sigma$ level relative to the expected dust contribution to the signal, and that the W2 signal in total is only a ~$4\sigma$ detection. With this in mind, a calculation can be made regarding the active area for the CO2. Using the values for Z, shown in log units of mol s$^{-1}$ m$^{-2}$ in Meech & Sorvan (2004), we compare our derived $Q_{CO2}$ and $Q_{CO}$ listed in Table 2 with the $Z_{CO2}$ $\sim 8 \times 10^{20}$ and $Z_{CO} \sim 1 \times 10^{22}$ mol s$^{-1}$ m$^{-2}$, which yields an active area of $\sim 10^5$ m$^2$ and

$4 \times 10^4$ m$^2$, or 0.3% to 0.1% of the nucleus surface area, for $CO_2$ and $CO$, respectively,

at a distance of 3.32 AU. Using Meech & Sorvan (2004) to scale the reported water production rates from Combi et al. (2010), and assuming constant active area, we find a $CO_2/H_2O$ production ratio $\sim 30\%$ at 3.32 AU, provided the W2 signal excess is from $CO_2$ emission.

## 4. Conclusions

This preliminary analysis of the WISE data for comet 67P/Churyumov-Gerasimenko

yields the following results:

- The nucleus size and presence of large-grain dust, and the rotation light
  curve amplitude values we derive here with the data from WISE are
  consistent with previously reported results in the literature.

- We provide new constraints on the age and nature of the dust, with mean
  grain size at least 1mm, as well as dust production rate estimates for the
  large-grain dust to be on the order of $10^4$ g/s.

- We find possible $CO/CO2$ emission at 3.32 AU, at production rates of 5 (+/-
  2) $\times 10^{25}$ mol s$^{-1}$ for $CO_2$, and no evidence at 4.18 AU down to an upper limit
  of $10^{26}$ mol s$^{-1}$.

- We provide 3$\sigma$ constraints on the albedo of the large-grain dust, finding a
  rough estimated for reflectance in the range of 0.05 +/-0.04, and a firm
  constraint for the reflectance at 3.4$\mu$m to be $\leq 0.12$. These reflectance
  constraints suggest the grains are not dominated by bright volatiles.

We speculate that the behavior of 67P may not be greatly unlike that of 103P, as the comet has many similar observed features, including a persistent large-grain dust trail and likely $CO_2$ production.

## 5. Acknowledgements

This publication makes use of data products from the Wide-field Infrared Survey Explore, which is a joint project of the University of California, Los Angeles, and the Jet Propulsion Laboratory/California Institute of Technology, funded by the National Aeronautics and Space Administration. This publication also makes use of data products from NEOWISE, which is a project of JPL/Caltech, funded by the Planetary Science Division of NASA. This material is based in part upon work supported by the NASA through the NASA Astrobiology Institute under Cooperative Agreement No. NNA09DA77A issued through the Office if Space Science. R. Stevenson is supported by the NASA Postdoctoral Program, and E. Kramer acknowledges her support through the JPL graduate internship program.

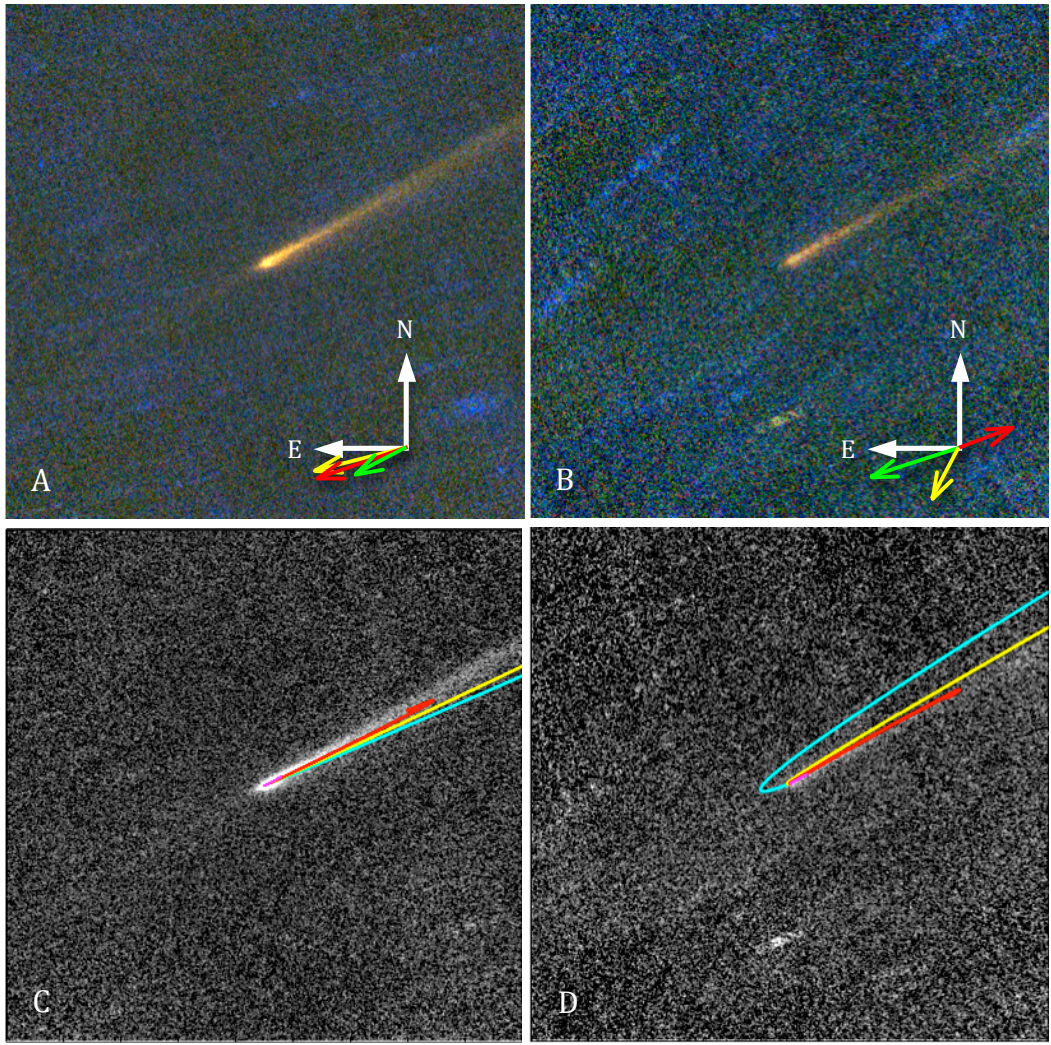

Figure 1: (A) Three-color image of comet 67P, half an arcminute on a side. The three longest wavelength bands, W2, W3, and W4, are mapped to the blue, green and red channels from the stack of images taken January 18-19, 2010. The projection vectors for sky-plane North and East (white), apparent motion (yellow), projected velocity vector (green), and sunward vector (red) are shown in the lower right corner. (B) Three-color image of comet 67P from June 29-30, 2010, with analogous insets as for (A). (C) Two-year syndynes overlaid on the January W3 image for a range of particle β values: 0.0001(magenta), 0.001(red), 0.01 (yellow), and 0.1(cyan). (D) 2-year syndynes for the June data with similarly colored syndynes. In both (C) and (D) the β=0.1 syndynes clearly extend beyond the observed coma, while the β=0.001 syndynes reside within the dust. Note also the hint of a residual trail preceding the comet in (C).

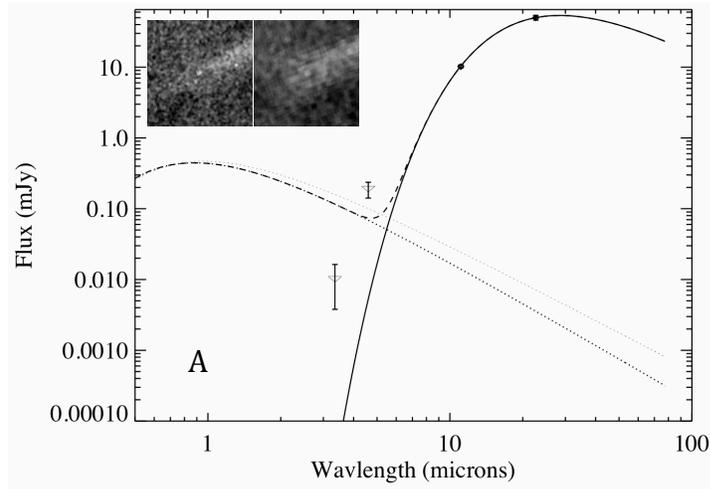

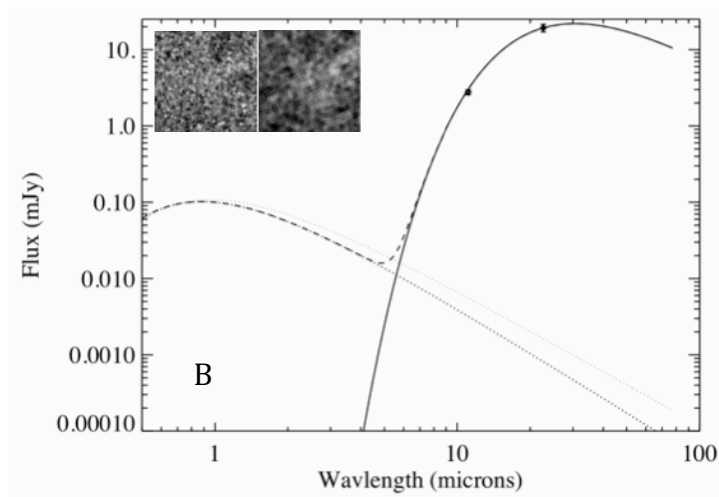

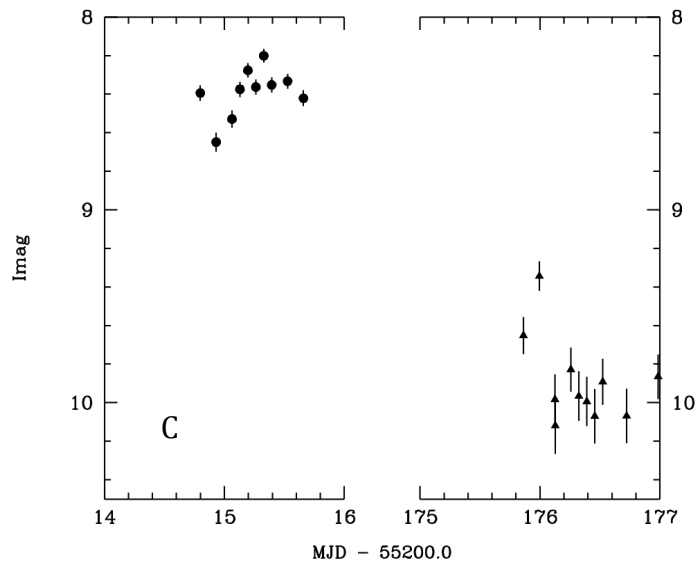

Figure 2: (A) Thermal black body fit (see Table 2) to the January 2010 photometry of the coma, and W2 infrared excess. The W3 & W4 extraction of the nucleus from the coma is shown in the upper left inset. (B) A similar fit for the June 2010 data. (C) The light curve in W3 instrumental magnitudes for January 18-19, 2010 and June 28-29, 2010.  The break in the horizontal axis represents the gap in the VA and VB data sets spanning Jan 20, 2010 – Jun 27, 2010.